# Self-Healing in Dielectric Capacitors: a Universal Method to Computationally Rate Newly Introduced Energy Storage Designs

Nadezhda A. Andreeva[1] and Vitaly V. Chaban[2]

(1) Peter the Great St. Petersburg Polytechnic University, Saint Petersburg, Russia. E-mail: andreeva_na@spbstu.ru.

(2) Yerevan State University, Yerevan, 0025, Armenia. E-mail: vvchaban@gmail.com.

**Abstract**.

Metal-film dielectric capacitors provide lump portions of energy on demand. While the capacities of various capacitor designs are comparable in magnitude, their stabilities make a difference. Dielectric breakdowns – micro-discharges – routinely occur in capacitors due to the inevitable presence of localized structure defects. The application of polymeric dielectric materials featuring flexible structures helps obtain more uniform insulating layers. At the modern technological level, it is impossible to completely avoid micro-discharges upon device exploitation. Every micro-discharge results in the formation of a soot channel, which is empirically known to exhibit semiconductor behavior. Because of its capability to conduct electricity, the emerged soot channels harm the subsequent capacitor performance and decrease the amount of stored energy. The accumulation of the soot throughout a dielectric capacitor ultimately results in irreversible overall failure. In the context of the dielectric breakdown, self-healing designates a range of chemical processes, which spontaneously rearrange the atoms in the soot channels to partially return their insulative function. We developed a universal method capable of rating new capacitor designs including electrode and polymer material and their proportions. We found the best-performing designs produce abundant volatile by-products after micro-discharge, whereas the soot samples exhibit lower electronic conductivities. We proved the capability of the theoretical method to rate the empirical performance of the known capacitors. The method relies on various electronic-structure simulations and potential landscape explorations. The predictions are validated through FTIR experiments. The reported advance opens an impressive avenue to computationally probe thousands of hypothetical capacitor designs.

**TOC Graphic**

The soot composition and microscopic structure influence self-healing efficiency in metal film capacitors.

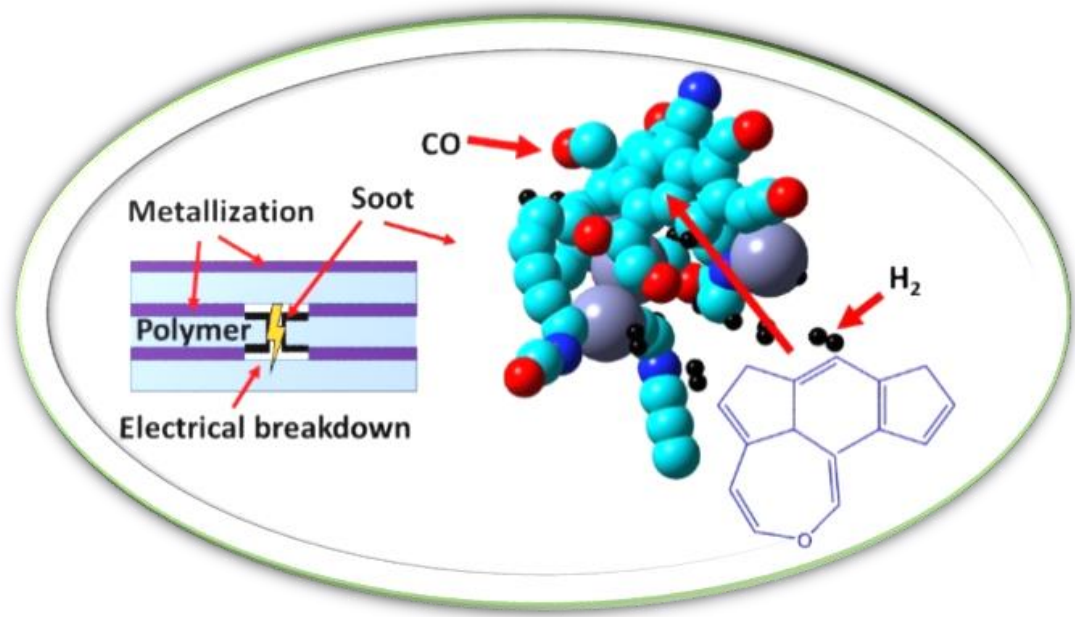

**Introduction**

The progress of civilization heavily depends on the availability of substantial amounts of energy to power novel technologies, such as supercomputing, ecologically friendly production of chemicals, thoughtful recycling of waste, and artificial intelligence training.[1-3] This demand concerns both sustainable energy sources and rechargeable energy storage.[4-6] The dielectric capacitors remain among the primary solutions to accumulate large portions of electrical energy. These devices are composed of electrodes made of metals or conductive polymers and dielectric polymers in between.[7-8] The spatial separation of the electrostatic charges stands beyond the capacitance of such a capacitor.[9]

The performance of dielectric capacitors hinges on a carefully orchestrated interplay between an electrode and a dielectric polymer. Choosing the right combination of materials determines not only charge storage capacity but also the reliability of functioning including the critical electrical breakdown threshold.[10] The ideal electrode material exhibits high conductivity, chemical stability in the operating environment, and – importantly – a good adhesion to the dielectric.[11] Higher-permittivity polymers like polyethylene terephthalate can store more charge. At the same time, these materials are more susceptible to the dielectric breakdown. Another popular dielectric polymer, polypropylene, exhibits an outstanding breakdown strength, making it a popular choice for high-voltage applications. Crystalline polymers offer better mechanical and thermal stabilities as compared to amorphous polymers.[12-15]

In the context of a capacitor's stability upon its exploitation, the interface between the electrode and dielectric is crucial.[16] Improper mutual adhesion is deemed to be harmful. It can lead to charge injection and localized electric field enhancements. These phenomena ultimately contribute to the dielectric breakdown. Surface modifications and adhesion promoters help to optimize the interface morphology and enhance capacitor performance.

Electrical breakdown in dielectric capacitors is a complex physical and chemical phenomenon driven by both intrinsic and extrinsic factors. The intrinsic breakdown involves mechanisms occurring within the bulk dielectric material, for instance, avalanche ionization. When electrons are accelerated by the electric field, they collide with molecules. Their collisions generate electron-hole pairs and foster a cascading ionization process that leads to current conduction. Additionally, the valence electrons may exercise tunneling applying their intrinsic wave functionalities. Electrons quantum mechanically penetrate the potential barrier between the localized states within the polymer, leading to sudden current flow. The extrinsic breakdown involves defects and inhomogeneities within the capacitor structure, including voids and inclusions. Gas-filled voids or conductive particles within the dielectric gradually concentrate the electric field and initiate localized breakdown. Moreover, ionic impurities within the polymer tend to migrate under the electric field. Such migrations create conductive paths and lower the theoretically computed breakdown voltage.

Understanding the dominant mechanism leading to dielectric breakdown for a specific combination of materials and operating conditions is crucial for improving the design and reliability of the engineered capacitor. One of the paramount things to understand in this context is manifold chemical transformations, which engage both electrode and polymer. The occured dielectric breakdown ruins the original materials by providing high-energy phonons giving rise to less organized chemical structures featuring different electrical and physicochemical properties. The compositions of the destroyed materials directly influence the identities of the dielectric breakdown products.[17]

The high temperature caused by the liberated energy of the dielectric breakdown event triggers a fascinating cascade of chemical transformations within the dielectric polymer. Near the breakdown site, a plethora of new compounds emerge. The latter affect the post-breakdown state and the subsequent capacitor's performance. This vast variety of compounds includes both metal atoms previously belonging to the electrode and non-metal atoms previously belonging to the dielectric polymer.

Previously we investigated the breakdown products of polypropylene, polyethylene terephthalate, polyethylene naphthalate, and polyphenylene sulfide using classical reactive molecular dynamics. We found that gaseous by-products constitute a significant fraction of the newly formed structures. Methane ($CH_4$), ethane ($C_2H_4$), carbon monoxide (CO), and molecular hydrogen ($H_2$) are commonly observed due to chain scission and fragmentation of the polymer chains. Low-molecular-weight oligomers and monomers also form.[18] Next, the formation of the conductive channels is realistic. High temperatures and disturbed electric fields carbonize the polymer in the vicinity of the breakdown site, creating conductive channels that perpetuate current flow. These reactions take place via chain scission, impacting the mechanical and electrical properties of the surrounding dielectric. Throughout our discussion, we designate all non-volatile compounds formed out of the electrode and the polymer as soot.

The specific compounds formed and their electrical properties depend on the breakdown mechanism, polymer type, and operating conditions. The investigation of these by-products provides valuable insights into the severity of the damage. It must guide strategies for material improvement and capacitor design. To design safer and more efficient energy storage systems for a multitude of applications, we herein investigate the emerged chemical compositions as a function of the dielectric polymer. We pay particular attention to the role of the electrode remainders (zinc atoms) and their interactions with the remainders of the polymers. The analysis of the relative energetics of the new molecules and their structures allows one to rate the trialed dielectric materials according to their capabilities to minimize the adverse consequences of the breakdown.

In this work, we calculated the electronic properties, such as conductivity, band gap, and density of states of the breakdown products for polypropylene (PP), polyethylene terephthalate (PET), polycarbonate (PC), and Kapton with zinc electrodes. The calculation showed that the breakdown products of Kapton have the highest conductivity, while those of PP have the lowest. Based on the results obtained, the polymers can be arranged in order of decreasing self-healing efficiency: PP > PET > PC > Kapton. The obtained trend is in agreement with real-world laboratory practice. PP is the most frequently dielectric in commercial devices, whereas PET is deemed to be its acceptable substitute. In turn, neither PC nor Kapton are employed in real modern dielectric capacitors. Whereas Kapton often gets praised for its unprecedentedly high resistance at the beginning of exploitation, the reliability of such capacitors is low. Our newly elaborated method explains the reasons beyond such a behavior and provides convincing molecular mechanisms of the damages.

**Methodology**

Multiple types of atomistic simulations were applied consequently to characterize the low-energy structures of the soot samples along with their thermodynamic and electronic properties: (1) potential energy surface explorations to reveal the most probable structural patterns based on non-periodic systems; (2) Hessian calculations to obtain vibrational spectra characterizing interatomic couplings; (3) density functional theory calculations with the periodic plane-wave basis and core potentials to evaluate the electronic properties of the soot samples with eliminated boundary effects. In the following, the due details of all performed simulations are provided.

The identification of the chemical identities of thermodynamically stable soot structures relies on the potential energy landscape exploration for given system compositions. Since the energy of the dielectric breakdown corresponds to subplasmic temperatures, it fosters the atomization of the electrode and surrounding polymer. At such temperatures, the diversity of the chemical reactions is not suppressed because the associated activation barriers can be swiftly overcome by all reacting supraatomic particles. Hence, the identities of the products are determined exclusively by the thermochemistry of each potential chemical transformation and the aggregate alteration of the system's thermodynamic potentials. Since breakdown-related chemical destruction takes place at high temperatures, the role of entropy is a cornerstone.

The potential energy surface was investigated according to the kinetic energy injection method.[19-21] The simulated systems were perturbed every 1,000 integration timesteps by the portion of kinetic energy corresponding to 1000, 2000, and 3000 K. The excessive energy was implemented by adding the Maxwell-Boltzmann distribution of translational momenta to every atom. After the energy injection, the system geometry was propagated via the conventional equilibrium PM7-MD simulation.[22-24] The excessive energy was gradually sucked away by the Berendsen thermostat,[25] whereas the potential energy evolved in line with the equipartition theorem. Before the next perturbation, the geometry of the system was recorded for further local energy minimization. This procedure corresponded to a complete removal of a kinetic energy effect on the stationary point geometry. The goal of the described procedure was to support the formation of various reasonable geometries, which should be expected to populate the soot structures.

The above-described procedure was repeated two hundred times for every chemical composition investigated. The energies of all detected molecular configurations were compared. The lowest-energy structure was proclaimed a system's global minimum. The equations of motion were propagated following the Verlet algorithm with a timestep of $1 \times 10^{-4}$ ps given high simulated temperatures. The temperature coupling constant was set accordingly to the employed timestep, to $1 \times 10^{-2}$ ps to preserve the numerical stability of the simulation.

The molecular configurations of particular interest (primarily, low-energy geometries) were additionally reoptimized using hybrid density functional theory (HDFT) to derive more accurate descriptors. The M11 exchange-correlation meta-GGA functional was used to approximate the wave equation.[26] M11 was trained against a large reference set of molecules. Furthermore, it adds semiempirically assessed dispersion (London) forces to every atom on the fly. M11 provides good accuracy of molecular geometries and thermodynamic properties. The collection of the electronic orbital functions proposed by Weigend and Ahlrichs[27] was selected to expand the wave function with the convergence criterion of the latter of $10^{-7}$ hartree. To calculate the mid-infrared and far-infrared vibrational spectra, the split-valence, double-zeta, polarized, atom-centered basis set 6-31G(d) for the involved second-row elements – C, H, O, N – was used. Since 6-31G(d) does not supply functions for Zn, the LANL2DZ basis set was chosen for this particular element. The numerical harmonic frequencies were computed for the reoptimized geometries of the soot samples. The vibrational FTIR spectra were accordingly predicted. The computed spectra can be directly compared to the experimental FTIR spectra for the soot samples. The partial atomic charges were evaluated to characterize the polarities of the systems. At this stage, non-periodic HDFT calculations were preferred over plane-wave periodic Kohn-Sham DFT calculations (KSDFT) because the former performs more accurately for the particular investigated properties. Partial electrostatic charges (ESP) were calculated using Merz-Kollman algorithm.[28]

The KSDFT calculations were employed to determine the band gap, density of states (DOS), and electrical conductivity. The Perdew–Burke–Ernzerhof (PBE) exchange-correlation functional was used in conjunction with the plane wave (PW) basis sets. The calculations were carried out using the open-source software Quantum Espresso 6.1 (QE).[29] The projector-augmented wave (PAW) method was used in lieu of core potentials. The electronic structure convergence criteria for the last change of total potential energy was set to $10^{-4}$ a.u. The force convergence criteria was set to $10^{-3}$ a.u. The number of k-points for all compositions was set to 27 to sample the Brillouin space. The plane-wave cut-off energy was set to 73 Ry. The results of the due benchmarking calculations are demonstrated in the Supplementary Information, Figures S1-S2.

The electrical conductivity under a direct current (DC) was calculated as the average trace using the Kubo-Greenwood theorem.[30] This type of calculation takes into account intra-band, inter-band, and degenerate-band conductivity contributions. It employs the Lorentzian representation of the Dirac delta function.[30]

Several general-purpose and in-house program codes were consequently used in this research project. MOPAC (version 22)[31] was employed to run millions of single-point PM7 semiempirical calculations.[32-34] The in-house codes implement the PM7-MD and global minimum search algorithms. GAMESS (version 2020)[35] was used to apply HDFT. The visualization of the atomistic structures and the preparation of graphics were carried out in VMD[36] and XCrySDen.[37] Matplotlib is a visualization library in Python that helped to construct some plots from the KSDFT calculations.[38] The computer code of KGEC was employed to get the reported electrical conductivities.[30]

**Results and Discussion**

To sample an entire configurational space, energy injections are of paramount importance. In this research project, we aimed to account for all possible chemical transformations to foster the formation of the lowest-energy polymer decomposition products, such as soot and gases. We probed a few degrees of perturbation to find a numerical balance between fast reactions and physical atomic movements. The usage of molecular dynamics to locate various stationary points assumes that the periodic perturbations should not totally suppress the Hamiltonian-based atomic displacements. It is essential to generate a realistic stationary point candidate before minimizing the involved interatomic forces. The speed of removal of the excess kinetic energy depends on the thermostat-related constants, which regulate coupling intensity to the temperature bath. The simulated soot was cooled from 1000 down to 400 K within less than 0.1 ps as a result of one of the benchmarked simulation setups (Figure 2). If any simulated system were unable to lose all of its excessive kinetic energy, the atomic momenta were zeroed and substituted by a newly generated Maxwell-Boltzmann distribution. We note that there is no need to reproduce the initial temperatures in the epicenter of the dielectric breakdown, assessed as being roughly 7000 K because such an energy also eliminates low-molecular products of the polymer decomposition. The emergence of the soot and the molecules of the by-products starts no earlier than the breakdown region cools below 3000 K.

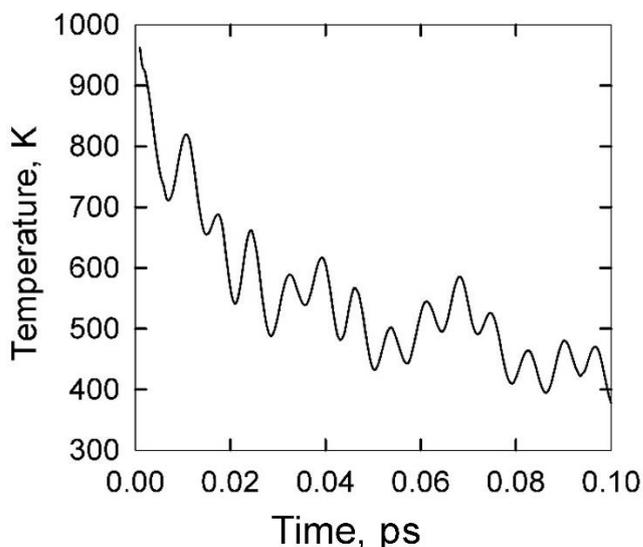

Figure 2. Temperature evolution in the 4 Zn + PP system after the kinetic energy injection. The thermostat reference temperature was set to 300 K in the present case.

As we have already discussed above, the efficiency of self-healing is a technological cornerstone determining the quality of a metal-film capacitor. The dielectric breakdown of the insulating film occurs near the location of the local defect. The breakdown core burns out. This small part of the capacitor, de facto, gets isolated. The spatial isolation allows the capacitor to continue operating without significant loss of its original capacity. New compounds formed during the breakdown make up the soot, which is semiconducting, and partially go to the gas phase, which is insulating.

The conductive behavior of the products formed during the breakdown process is affected by the original composition of the polymer and capacitor plates. In the present work, we selected four very common dielectrics used in modern dielectric capacitors. Their chemical structures are summarized in Figure 2 to imagine elemental compositions.

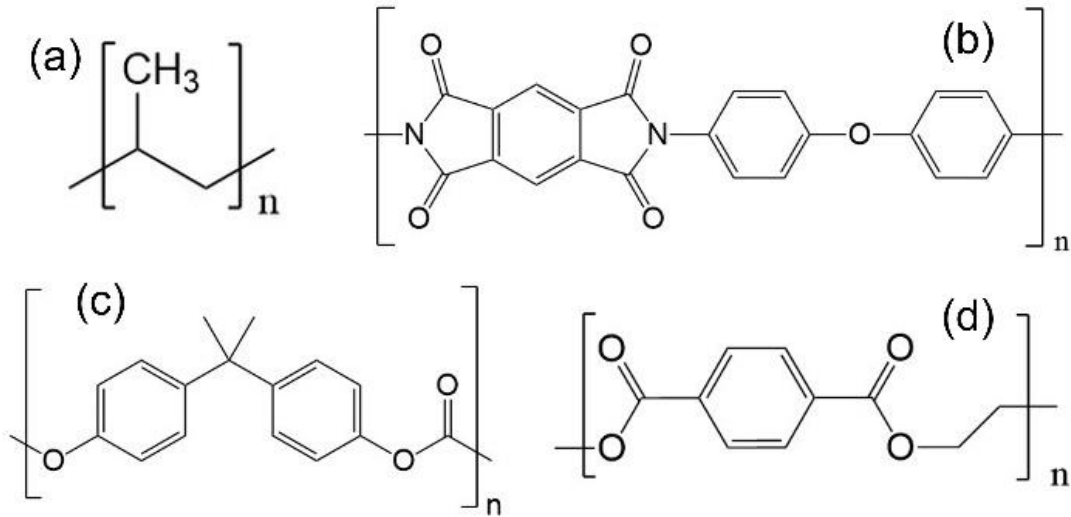

Figure 2. The polymers investigated in the present work are (a) polypropylene, (b) poly-oxydiphenylene-pyromellitimide better known as Kapton, (c) polycarbonate, and (d) polyethylene terephthalate.

With a rough thickness ratio of electrode/polymer equaling 1/50, the ratio of zinc to polymer atoms can be approximated to be near 1/100. A straightforward estimation of a trustworthy model composition was performed using the following formula:

$$\frac{n(polymer)}{n(Zn)} = \frac{d_{pol} h_{pol} M_{Zn}}{M_{pol} d_{Zn} h_{Zn}},$$

where $n(polymer)$ and $n(Zn)$ are amounts of substances, $d_{pol}$ and $d_{Zn}$ are specific densities, $h_{pol}$ and $h_{Zn}$ are thicknesses, $M_{pol}$ and $M_{Zn}$, represent molar masses of a polymer and zinc, respectively.

We need to take into account the following empirical fact. When a part of the capacitor burns out, the area of the burnt-out electrode is always greater than the area of the destroyed polymer. Thus, we investigated systems, in which the number of zinc atoms was slightly greater than in the real case of breakdown. In addition, there is a long-standing interest in measuring the impact of an increased metal atom content on the electrical conductivity of the soot sample. Such knowledge may provide an essential aid in thoughtfully adjusting the proportions of the capacitor constituents to obtain unusual macroscopic effects. Apart from the investigation of the polymer performance, we characterized the effect of the zinc atom fraction on the metal-nonmetal chemical bonding, band gap, and conductivity alterations.

To investigate the properties of the semiconducting soot, nine atomistic systems were created. Their descriptive features are presented in Table 1. Since a dielectric breakdown leads to a full atomization of a system, the initial geometries taken for simulations do not make any actual sense. For each chemical composition, 200 iterations of the kinetic energy injection method were carried out providing, accordingly, 200 non-unique stationary point configurations (Figure 3).

Table 1. Fundamental details describing the simulated chemical compositions.

| # | Composition | # nuclei | # electrons | Abbreviation |
|---|---|---|---|---|
| 1 | $[C_3H_6]_{10}$ | 90 | 240 | PP |
| 2 | $2\,Zn + [C_3H_6]_{10}$ | 92 | 300 | 2 Zn + PP |
| 3 | $4\,Zn + [C_3H_6]_{10}$ | 94 | 360 | 4 Zn + PP |
| 4 | $6\,Zn + [C_3H_6]_{10}$ | 96 | 420 | 6 Zn + PP |
| 5 | $8\,Zn + [C_3H_6]_{10}$ | 98 | 480 | 8 Zn + PP |
| 6 | $10\,Zn + [C_3H_6]_{10}$ | 100 | 540 | 10 Zn + PP |
| 7 | $4\,Zn + [C_{22}H_{10}O_5N_2]_2$ | 82 | 512 | 4 Zn + Kapton |
| 8 | $4\,Zn + [C_{16}H_{14}O_3]_3$ | 103 | 522 | 4 Zn + PC |
| 9 | $4\,Zn + [C_{10}H_8O_4]_5$ | 114 | 620 | 4 Zn + PET |

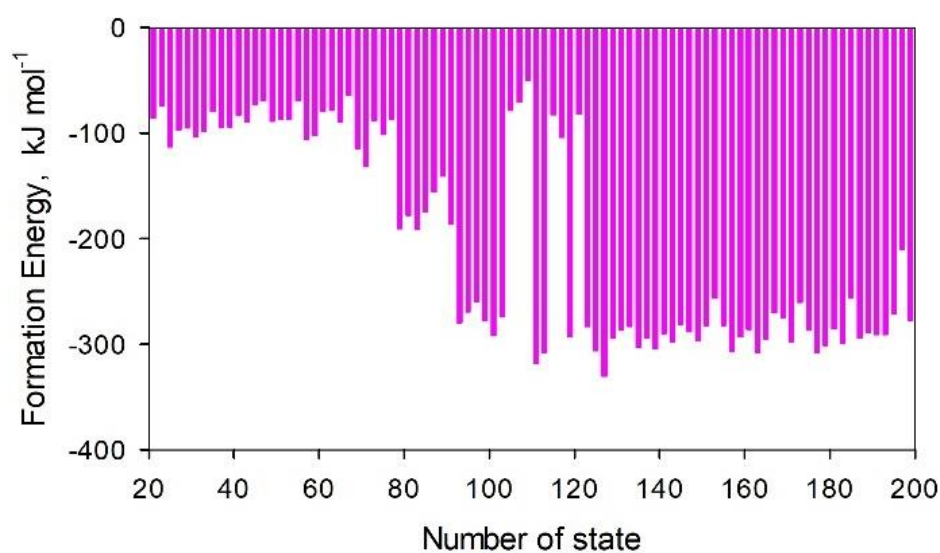

Figure 3. The exemplary distribution of the stationary point formation energies for the [4 Zn + PP] composition. The first nineteen recorded stationary points were disregarded since their formation energies appeared to be very high. The thermodynamic stabilities of such chemical structures are marginal.

The global minimum search using the kinetic energy injection method was started from a nearly random arrangement of atoms taken in the proportion characteristic of each precursor polymer. Therefore, the energies of the first detected stationary points were particularly high being expectedly far from the low-energy region of the phase space. The kinetic energy injection method makes a simultaneous use of molecular dynamics which drifts along the gradient of the descending potential energy and abrupt periodic random energy injections to ignore the activation barriers on the way. Consequently, we obtained a group of generally higher-energy stationary points coming first and a group of generally lower-energy ones coming up to the middle of the plot and persisting until its end. Figure 3 convincingly demonstrates this feature of the employed method.

The dispersion of the dataset in Figure 3 amounts to 95.5 units excluding 19 most energetic stationary points, e.g., the ones with a formation energy of over -100 kJ/mol. The population of such geometrical configurations in a real system is marginal and they can be safely excluded from the analysis of the electrical properties of the soot. Instead, 10 low-energy atomic configurations were selected, whose energies differed from the global one by less than the 3 kT product of room conditions.

The difference between the global minimum arrangement of atoms and the highest energy was found to equal 284 kJ/mol. The reported difference does not include transition states, if any, and the first high-energy configurations discussed above. The maximum energetic difference between the stationary points characterizes the flexibility of a given collection of atoms. In other words, this descriptor reflects the extent of the ability of a system to repack into various chemical structures exhibiting substantially different structural energies. The information may represent an interest in creating very sophisticated molecular designs.

The system [4 Zn + PP] composition of a dielectric capacitor, in its global minimum configuration of a soot species, produces 16 $H_2$ molecules, 2 $C_2H_2$ molecules, and a soot prototype molecule without any symmetric structure. No hydrocarbons have been detected as well as in higher-energy configurations (Figure 4a. The latter corresponds to an energy of -599 kJ mol$^{-1}$. In some configurations, a single cyclopropene molecule is present, Figure 4b, corresponding to an energy of –534 kJ mol$^{-1}$. On a related note, the metal atoms do not form covalent bonds with each other, preferring to form chains of the [–Zn–H…Zn–] type. Herein, the covalent bonding is indicated by a line and the electrostatic bonding is indicated by dots. Of four zinc atoms in many local minima points, two atoms form the above chain. The remaining zinc atoms are incorporated into the carbon chain. The Zn–H distance is 153 pm, which corresponds to a covalent bond based on the analysis of the available covalent radii. The H...Zn distance is 187 pm corresponding to an ionic bond. The distance between the zinc atoms is 300 pm, the Zn–H...Zn angle is 124 degrees. The Zn…C bond lengths are in the range of 199 – 211 pm. This data corresponds to ionic bonds. The angle C…Zn…C amounts to 150 – 159 degrees.

According to the partial charge analysis, Zn exhibits the strongest electrophilic properties in the soot samples. For instance, the Marz-Kollmann charges of Zn in [–Zn–H…Zn–] fragments are +0.85e to +1.03e. The most nucleophilic interaction centers are carbons, see [–C…Zn…C–] chains. The partial electrostatic charges assigned to the carbon atoms are -1.232e, -1.001e, -0.730e, and -0.693e.

The system [4 Zn + PP] contains the largest mass fraction of the gas phase, 12.3 wt.%, compared to all other systems studied in this work. Since many atoms of matter leave the prospective soot sample in the form of gas, they decrease the mass and volume of the soot. Smaller soot species naturally exhibit proportionally decreased macroscopic conductivities.

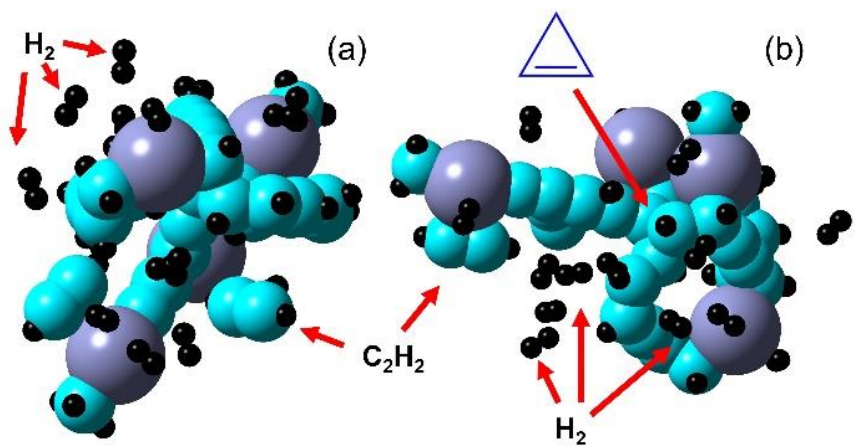

Figure 4. The geometries characterizing the [4 Zn + PP] system: (a) the structure of the global minimum; (b) the structure of one of the low-energy local minima. In both configurations, two molecules of $C_2H_2$ and 16 molecules of $H_2$ formed. The local minimum structure contains a single cyclopropene molecule. The carbon atoms are cyan, the zinc atoms are violet, and the hydrogen atoms are black.

When the number of zinc atoms in the system is increased by a factor of 2.5, cyclic hydrocarbons emerge in most of the resulting configurations. The most common is cyclopropene, as shown in Figure 5a, corresponding to the global minimum with an energy of -729 kJ mol$^{-1}$. In addition, other heterocyclic compounds also form, Figure 5b. They correspond to local minima with an energy of -522 kJ mol$^{-1}$. Zinc atoms prefer to connect indirectly but through a hydrogen atom. Of the 10 zinc atoms, nine are involved in the formation of bonds, [–Zn–H...Zn–], and even a [-Zn–H...Zn–H...Zn–] chain. This is because one covalent bond and one ionic bond involving Zn and H are more energetically beneficial than other spatial combinations. Indeed, the Zn–H distances range from 146 to 153 pm. These distances correspond to covalent bonds. The H...Zn distances are 179–196 pm corresponding to ionic bonds. The distances between the adjacent Zn atoms are 312-321 pm, and the Zn–H...Zn angles are 120-131 degrees. One zinc atom is incorporated into the carbon chain, forming a structure like [–C…Zn…C–]. The Zn…C distances are 203-206 pm indicating strongly polar covalent bonds. In turn, the C…Zn…C angles are around 152 degrees. The gas phase is formed by 13 $H_2$ molecules, which is 2.4 wt.%. No acetylene molecules exist.

In the [–Zn–H…Zn–] chains, the electrostatic charges of Zn vary from +0.669e to +0.910e. The charges of H vary from -0.480e to -0.325e. In the [–C…Zn…C–] chain, the charges of C are –0.540e and –0.347e, whereas of Zn is +0.788e. The electron-deficient Zn atoms attract the electron-poor C and H atoms belonging to the soot. Note that hydrogen atoms are nucleophilic reminiscent of those in the hydride structures.

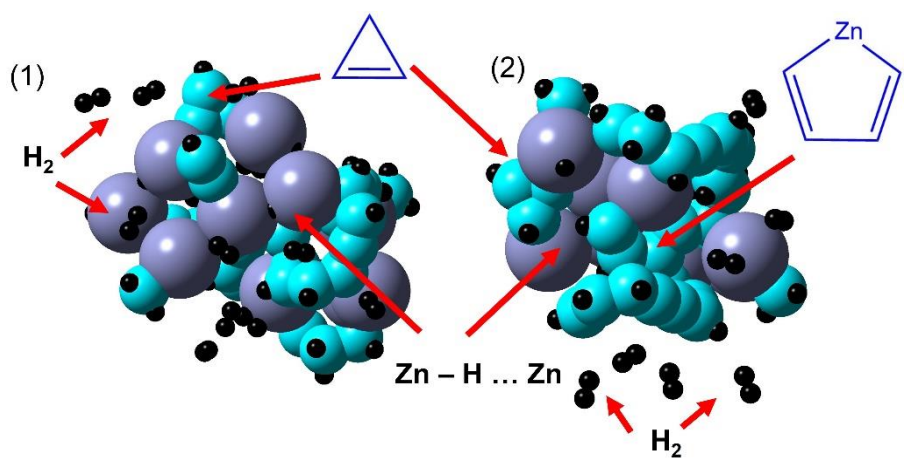

Figure 5. The geometries characterizing the [10 Zn + PP] system: (a) the structure of the global minimum; (b) the structure of one of the low-energy local minima. In both configurations, 13 molecules of $H_2$ formed. The structures contain depicted aromatic carbon-based rings. The carbon atoms are cyan, the zinc atoms are violet, and the hydrogen atoms are black.

Cyclic compounds are observed in most of the probable structures of the [4 Zn + PET] composition. Figure 6a displays the global minimum configuration of an energy –2937 kJ mol$^{-1}$. Unit structures include the cycles depicted in Figure 6b, a local minimum with an energy of -2799 kJ mol$^{-1}$. Ten molecules of hydrogen and 1-2 molecules of carbon monoxide form depending on a particular local minimum energy. In the resulting soot prototype molecule, zinc prefers to form ionic bonds with oxygen. Per six [-Zn…O-] bonds, two [-Zn…C-] bonds form. The ionic bond lengths between zinc and oxygen are 189-198 pm. The bonds between zinc and carbon appear to be 212-218 pm long. The O…Zn…O angles are either 90 or 135 degrees. The O…Zn…C angles are 101 and 136 degrees. The chains of the type [-Zn–H...Zn-] dot not form. The mass fraction of the gas phase in [4 Zn + PET] is 3.9-6.2 wt.%.

In [-C…Zn…O-], the electrostatic charges of carbon equal -1.009e and -1.172e. The charges of zinc are +1.199e and +1.127e. The charges of oxygen are -0.729e and -0.650e. In [-O…Zn…O-], the charges of oxygen are -0.677e and -0.771e. The charge of Zn is +1.288e. In [-C…Zn…C-], the charges of carbon are -0.895e and -1.482e, whereas the charge of Zn amounts to +1.278e.

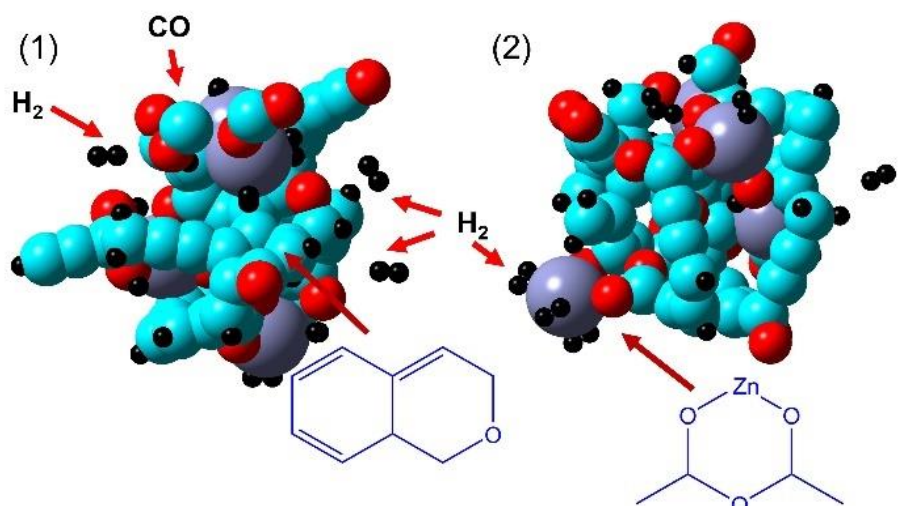

Figure 6. The geometries characterizing the [4 Zn + PET] system: (a)the structure of the global minimum; (b) the structure of one of the low-energy local minima. In both configurations, 10 molecules of $H_2$ formed. One molecule of CO in the global minimum structure and two CO molecules in the chosen local minimum structure emerged. The carbon atoms are cyan, the oxygen atoms are red, the zinc atoms are violet, and the hydrogen atoms are black.

In the [4 Zn + PC] system, the gas phase contains 6 hydrogen molecules, 1 carbon monoxide molecule, and 1 formaldehyde molecule, in all configurations. The soot prototype is partially represented by heterocycles and a complex structure of interconnected cyclocarbons as we observe in Figure 7a. It corresponds to the global minimum with an energy of –530 kJ mol$^{-1}$ and in Figure 7b, which corresponds to the local minimum with an energy of -370 kJ mol$^{-1}$. Due to the smaller number of oxygen atoms per modeled system (9 in PC versus 20 in PET), zinc forms only 2 bonds with oxygen. The lengths of Zn–O covalent bonds are 178–179 pm. The Zn…C ionic bond lengths are 198–205 pm. Two of the four zinc atoms form a chain of the form [-Zn–H...Zn-]. The Zn–H distance is 158 pm, H...Zn is 169 pm, the distance between zincs is 298 pm, and the Zn–H...Zn angle is 131 degrees. Mass fraction of gas phase 6.4 wt.%.

In the [-Zn–H…Zn-] chain, the electrostatic charges of Zn are +0.988e and +0.430e. The charge of H is –0.424e. In [-C…Zn–H), the charge of C is -0.848e, and the charge of Zn is +0.783e. The charge of H is -0.424e. In [-C…Zn…O-], the charge of C is –0.244e. Compare it to the cases of Zn and O, +0.443e and –0.339e, respectively.

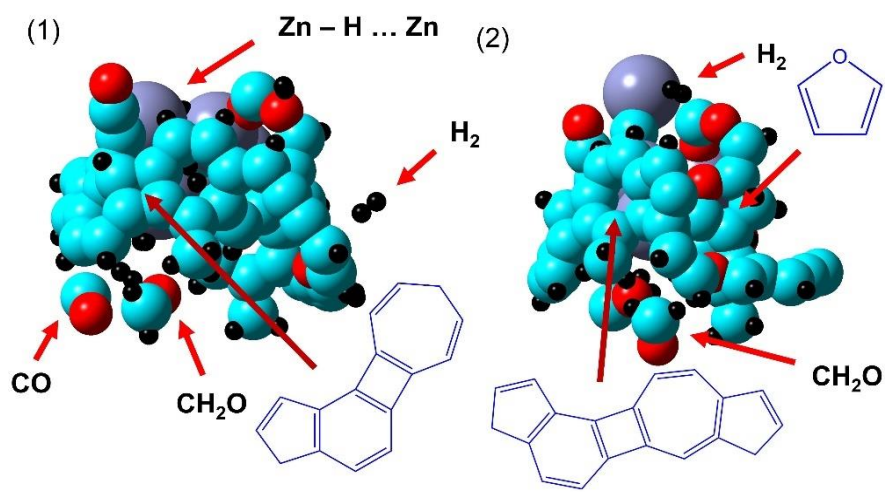

Figure 7. The geometries characterizing the [4 Zn + PC] system: (a) the structure of global minimum; (b) the structure of one of the chosen low-energy local minima. In both exemplified configurations, six molecules of $H_2$, one molecule of CO, and one molecule of formaldehyde formed. The depicted stationary point structures contain noteworthy carbon-based rings. The carbon atoms are cyan, the oxygen atoms are red, the zinc atoms are violet, and the hydrogen atoms are black.

The system [4 Zn + Kapton] shows a large variety of structures containing various nanocarbon species and heterocycles. The configuration of the global minimum is shown in Figure 8a with an energy of 647 kJ mol$^{-1}$ and one of the local minima in Figure 8b with an energy of 894 kJ mol$^{-1}$. The number of molecules in the gas phase differs depending on the configuration. The global minimum system contains 3 $H_2$ molecules, one CO molecule, and one $H_2O$ molecule. The local minimum configurations contain structures close to graphene-like, Figure 8b. The [-Zn-H...Zn-] chains are not formed. The distances between Zn…O are 189-195 pm and the distances between Zn…C are 199-205 pm correspond to ionic bonds. The lengths of Zn-H covalent bonds are 143-144 pm. The mass fraction of the gas phase amounts to 5.1 wt.%.

In [-C…Zn–H], the electrostatic charges of the carbon atoms are -0.712e and -0.686e. The charges of Zn are +0.909e and +0.714e. The charges of hydrogen atoms are -0.443e and -0.375e. In [-C…Zn…O-], the charges of carbon atoms are -0.341e and -1.062e. The charges of Zn are +0.767e and +0.960e. The charges of the oxygen atoms are -0.542e and -0.650e.

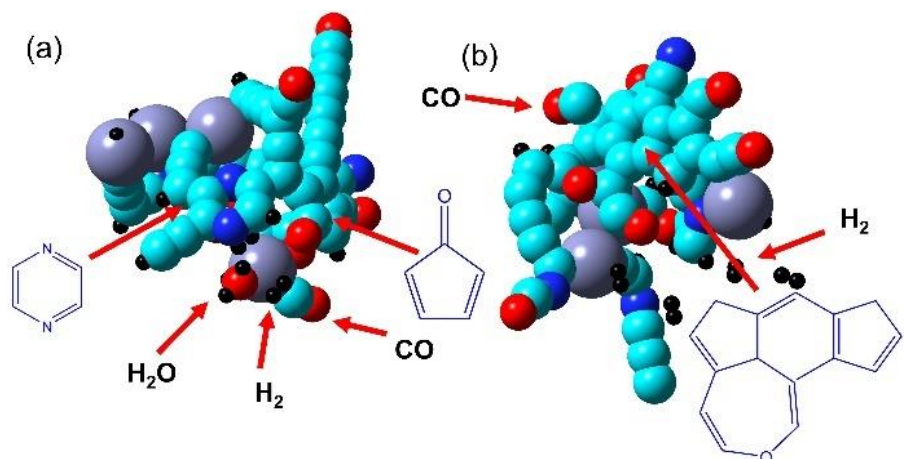

Figure 8. The geometries characterizing the [4 Zn + Kapton] system: (a) the structure of the global minimum; (b) the structure of one of the low-energy local minima. Three molecules of $H_2$, one molecule of CO, and one molecule of water formed in the global minimum configuration. In turn, the lower-energy configuration features four molecules of $H_2$ and one molecule of CO. The structures contain carbon, nitrogen, and oxygen-containing rings. The carbon atoms are cyan, the oxygen atoms are red, the nitrogen atoms are blue, the zinc atoms are violet, and the hydrogen atoms are black.

The configuration corresponding to the global minimum was further optimized using HDFT after removing molecules corresponding to the gas phase. Mid- and far-infrared vibrational spectra were calculated for the obtained structures at the hybrid DFT, M11/6-31G(d) + LANL2DZ (for Zn) level of theory.

Covalent and non-covalent infrared longitudinal wagging ($\omega$) vibrations between Zn and H are observed at 404 and 456 cm$^{-1}$ in the [4 Zn + PP] system, Figure 9. Symmetric stretching ($V_s$) vibration in Zn–H is recorded at 1851 cm$^{-1}$.

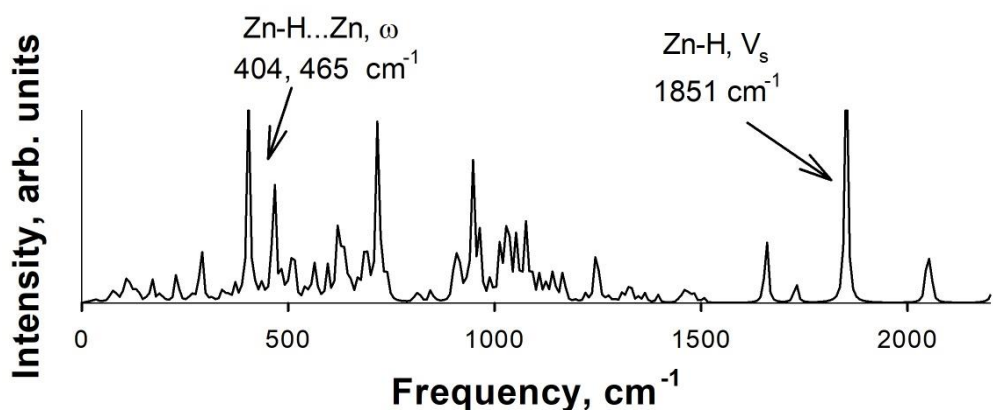

Figure 9. The computational mid-infrared and far-infrared vibrational spectra were recorded in the [4 Zn + PP] system.

Figure 10 shows the mid-infrared and far-infrared vibrational spectrum recorded in the [10 Zn + PP] system. An increase in the number of zinc atoms in the system leads to the appearance of new vibration peaks. The peaks corresponding to the latitudinal rocking ($\rho$) of the Zn–H bond are located at frequencies of 396-414 cm$^{-1}$. The peak characteristic of vibrations between covalent and non-covalent interactions of zinc and hydrogen Zn–H...Zn longitudinal wagging ($\omega$) corresponds to 497 cm$^{-1}$. Symmetric stretching ($V_s$) of Zn–H is observed at frequencies of 1659, 1865, 1924, and 1929 cm$^{-1}$. Antisymmetric stretching ($V_{as}$) for Zn–H...Zn is at a frequency of 1746 cm$^{-1}$.

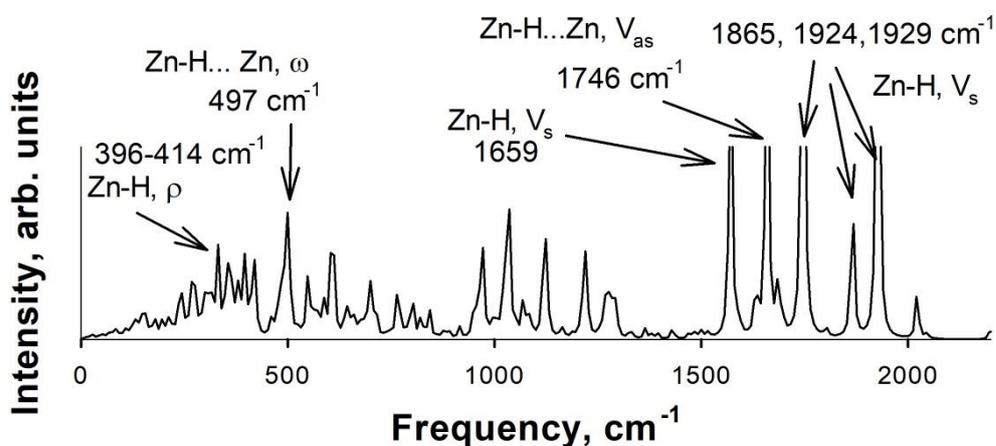

Figure 10. The computational mid-infrared and far-infrared vibrational spectra were recorded in the [10 Zn + PP] system.

In Figure 11, the mid-infrared and far-infrared vibrational spectrum correspond to the [4 Zn + PET] system. Vibrations corresponding to latitudinal scissoring (δ) occur at a frequency of 212 cm$^{-1}$, symmetric stretching (V$_s$) Zn…O is observed at a frequency of 308 and 318 cm$^{-1}$, antisymmetric stretching (V$_{as}$) Zn…O at 471 cm$^{-1}$.

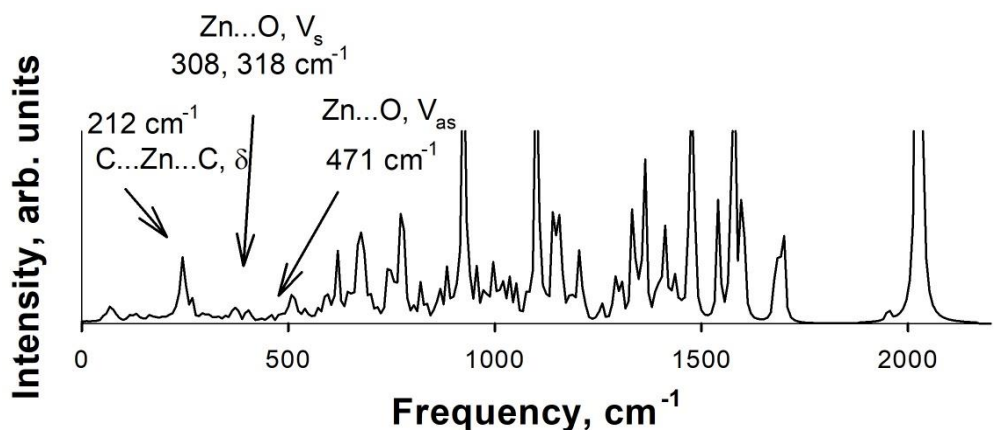

Figure 11. The computational mid-infrared and far-infrared vibrational spectra were recorded in the [4 Zn + PET] system.

In Figure 12, mid-infrared and far-infrared vibrational spectra were recorded in the [4 Zn + PC] system. The peak corresponds to a latitudinal scissoring (δ) of the C…Zn–H bond is located at a frequency of 397 cm$^{-1}$. The peak of longitudinal wagging vibrations (ω) for Zn–H...Zn is located at a frequency of 1058 cm$^{-1}$. The peaks characteristic of radial vibrations between covalent and non-covalent interactions of zinc and hydrogen Zn–H...Zn symmetric stretching (V$_s$) and antisymmetric stretching (V$_{as}$) are located at frequencies of 1058 and 1641–1653 cm$^{-1}$, respectively. Symmetric stretching (V$_s$) C…Zn–H is observed at a frequency of 1884 cm$^{-1}$.

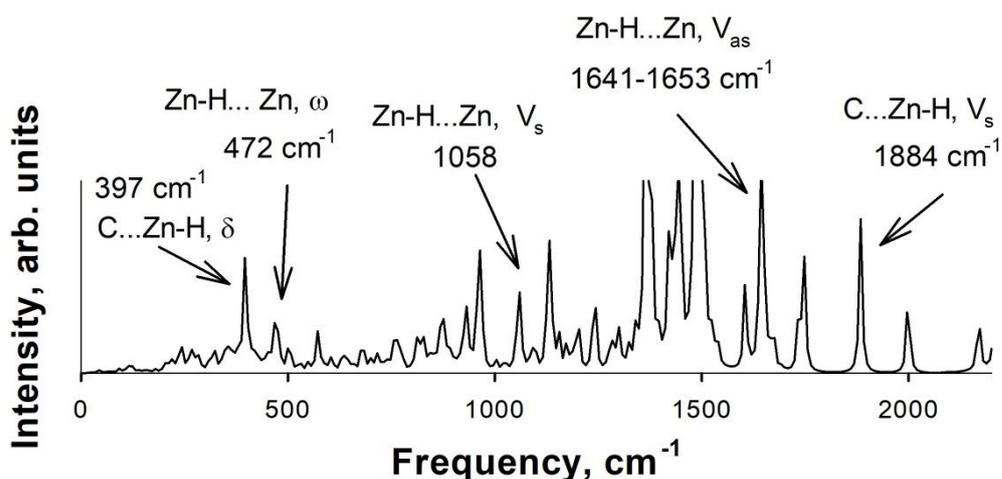

Figure 12. The computational mid-infrared and far-infrared vibrational spectra were recorded in the [4 Zn + PC] system.

In Figure 13, the mid-infrared and far-infrared vibrational spectra were recorded in the [4 Zn + Kapton] system. Peaks characteristic of this system: latitudinal rocking ($\rho$) for C…Zn–H at 330 cm$^{-1}$, latitudinal scissoring ($\delta$) of the C…Zn–H at 456 cm$^{-1}$, symmetric stretching (Vs) in the heterocycle for C–N–C is observed at 1227 cm$^{-1}$.

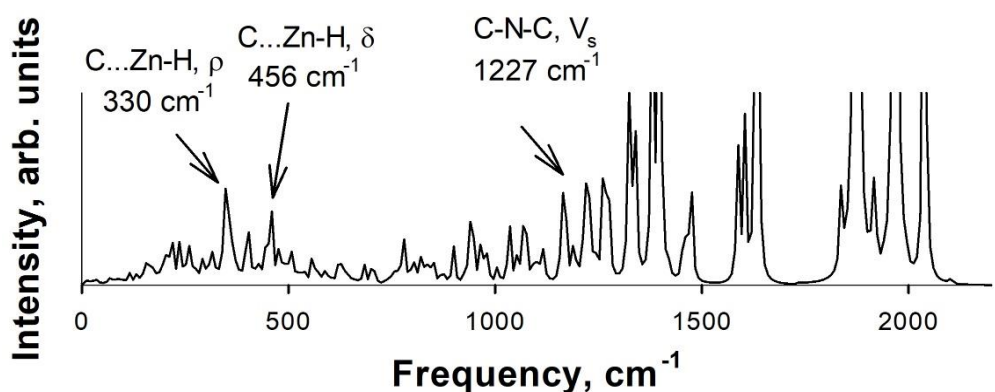

Figure 13. The computational mid-infrared and far-infrared vibrational spectra were recorded in the [4 Zn + Kapton] system.

The computed infrared spectroscopic signals can be used to verify the identities of the emerged bonded species in the soot vs. the experimental spectra of the in-lab soot samples. In such a way, key structural patterns representing the soot can be rationalized. The correspondence between the experimental and theoretical descriptions can be established. In the present work, we disregarded high-energy configurations, as rare ones, from the subsequent analysis. Only the configurations, whose energies were above the global minimum structure (for each composition) by less than 3kT, were investigated via KSDFT subject to gaseous by-products removal. The boundary effects were removed by reoptimizing the cell vectors to fit the cell content. Figure 14 shows fragments of periodic structures consisting of eight elementary cells.

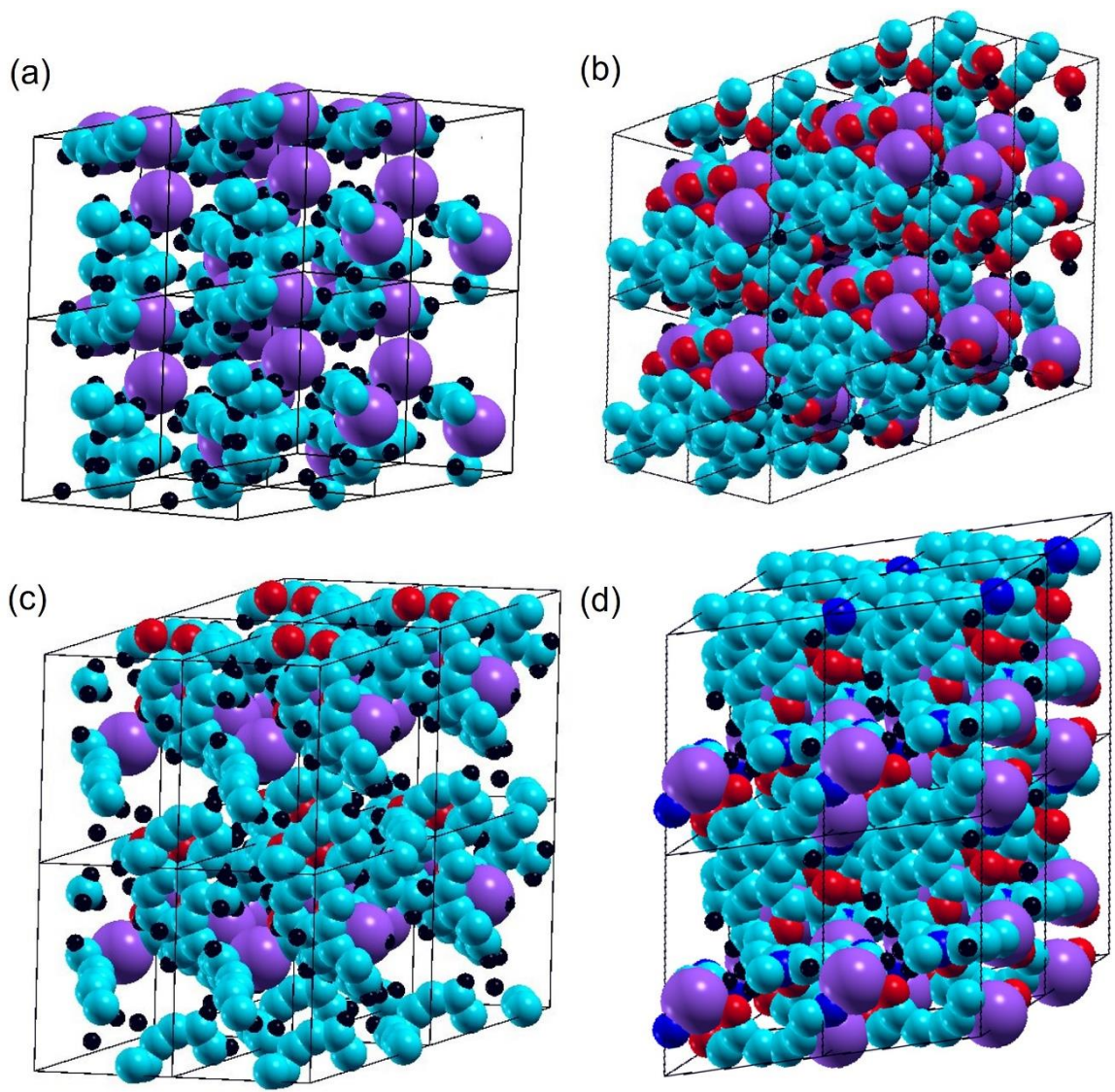

Figure 14. The fragments of the periodic simulation cells: (a) [4 Zn + PP] system; (b) [4 Zn + PET] system; (c) [4 Zn + PC] system; (d) [4 Zn + Kapton] system. Prior to the KSDFT simulations, all gas molecules were deleted. In turn, the unit cell volumes were reoptimized accordingly. The carbon atoms are cyan, the oxygen atoms are red, the nitrogen atoms are blue, the zinc atoms are violet, and the hydrogen atoms are black.

Self-healing is a group of spontaneous physicochemical processes, which partially eliminate the adverse effects of the dielectric breakdown by substituting a destroyed dielectric polymer with a semiconductor (soot). While such a substitute of materials does not represent full reversible healing, it prolongs the life cycle of a capacitor. A more efficient healing takes place if the soot sample exhibits as low electrical conductivity as possible. Using our methodology, it is possible to predict, which polymers give rise to a less-conductive soot.

The band gap and density of electronic states are descriptors that characterize the electrical conductivity of a material. Figure 15 shows the band structure (left) and DOS (right) of the [4 Zn + PP] system. Below the Fermi level, 2.6 eV, there is a valence band. Above the Fermi level, there is a conduction band. The difference between these energies is 1.7 eV. In DOS, the shaded areas indicate electron-filled levels, whereas the white areas indicate unfilled levels.

The maximum DOS value coincides with the HOMO level of the PP-based soot and is 26.7 eV$^{-1}$ a.u.$^{-3}$ at 1.69 eV. The DOS of the LUMO is 24.8 eV$^{-1}$ a.u.$^{-3}$ at 3.4 eV. The HOMO-1 level is located very close to HOMO, with a density of 26.7 eV$^{-1}$ a.u.$^{-3}$ HOMO-2, HOMO-3, and HOMO-4 merge into a single line. Their densities appear to be even higher, up to 55.1 eV$^{-1}$ a.u.$^{-3}$ However, the latter are separated from HOMO and HOMO-1 by ~0.2 eV.

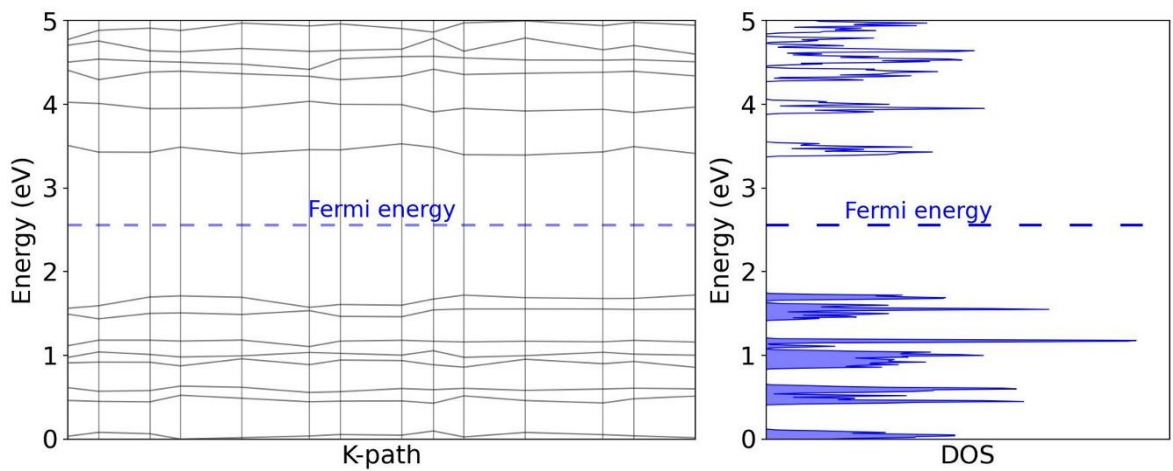

Figure 15. The band structure (left) and DOS (right) of the [4 Zn + PP] system. The band gap amounts to 1.7 eV. The Fermi energy equals 2.6 eV.

Figure 16 shows the band structure and DOS of the [4 Zn + PET] system. The fermi level is 2.9 eV, whereas the band gap is 1.7 eV. The maximum DOS value amounts to 86.8 eV$^{-1}$ a.u.$^{-3}$ The DOS of the LUMO is 90.3 eV$^{-1}$ a.u.$^{-3}$ at 3.2 eV. The HOMO-1 level is next to the HOMO. The corresponding density of states equals 43.3 eV$^{-1}$ a.u.$^{-3}$ The HOMO-2 level is located 0.09 eV below, with a density of states of 86.8 eV$^{-1}$ a.u.$^{-3}$

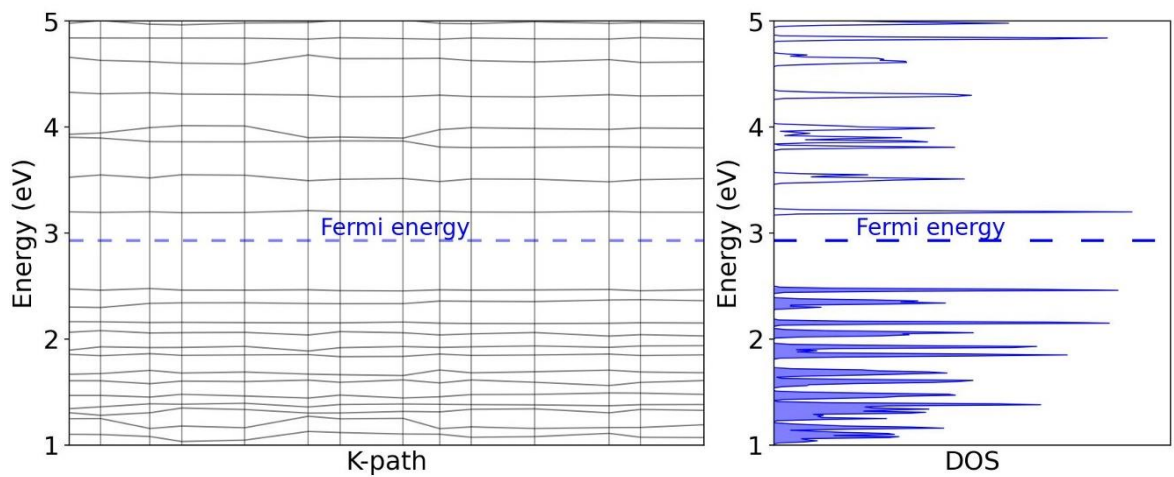

Figure 16. The band structure (left) and DOS (right) of the [4 Zn + PET] system. The band gap amounts to 0.7 eV. The Fermi energy equals 2.9 eV.

Figure 17 shows the band structure and DOS of the [4 Zn + PC] system. The Fermi level is 2.2 eV. The band gap is 0.6 eV. The HOMO level features the density of states of 42.7 eV$^{-1}$ a.u.$^{-3}$ The DOS of the LUMO is 23.7 eV$^{-1}$ a.u.$^{-3}$ at 2.3 eV. HOMO-1, HOMO-2, and HOMO-3 merge forming a line with a width of 0.29 eV. Its density of states amounts to 63.0 eV$^{-1}$ a.u.$^{-3}$ The energy gap between HOMO and the described line is 0.46 eV.

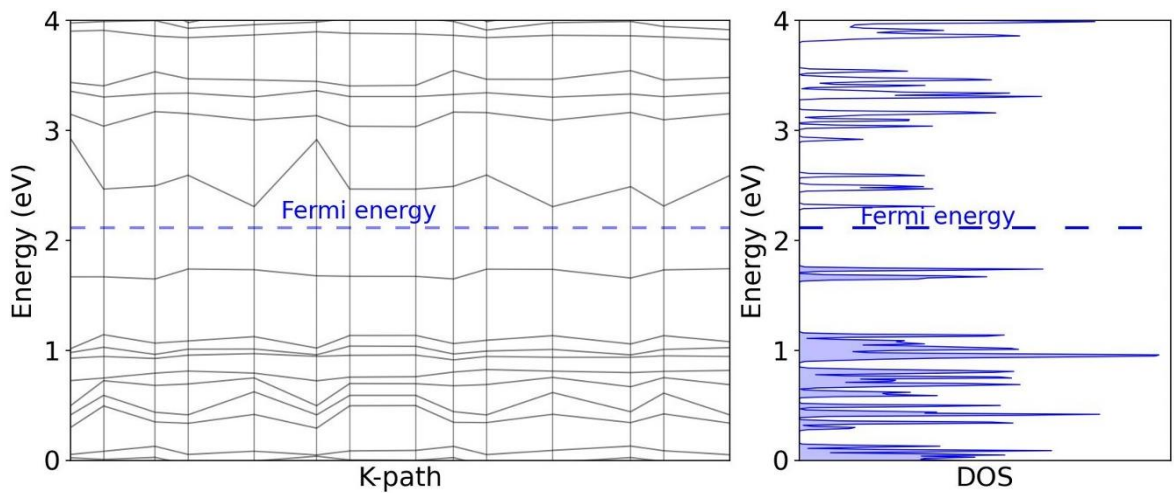

Figure 17. The band structure (left) and DOS (right) of the [4 Zn + PC] system. The band gap is 0.6 eV. The Fermi energy is 2.2 eV.

In the [4 Zn + Kapton] system, the valence electronic levels are located near the Fermi energy level, as well as the conduction band levels, although they do not cross it. Figure 18 shows the band structure (left) and DOS (right) of the [4 Zn + Kapton] system. The Fermi level equals 1.9 eV, whereas the band gap is 0.12 eV. The HOMO, HOMO-1, and HOMO-2 levels merge into a line, with a width of 0.5 eV, the maximum density of states amounting to 38.7 eV$^{-1}$ a.u.$^{-3}$ at 1.78 eV. The DOS of the LUMO is 61.4 eV$^{-1}$ a.u.$^{-3}$ at 2.2 eV.

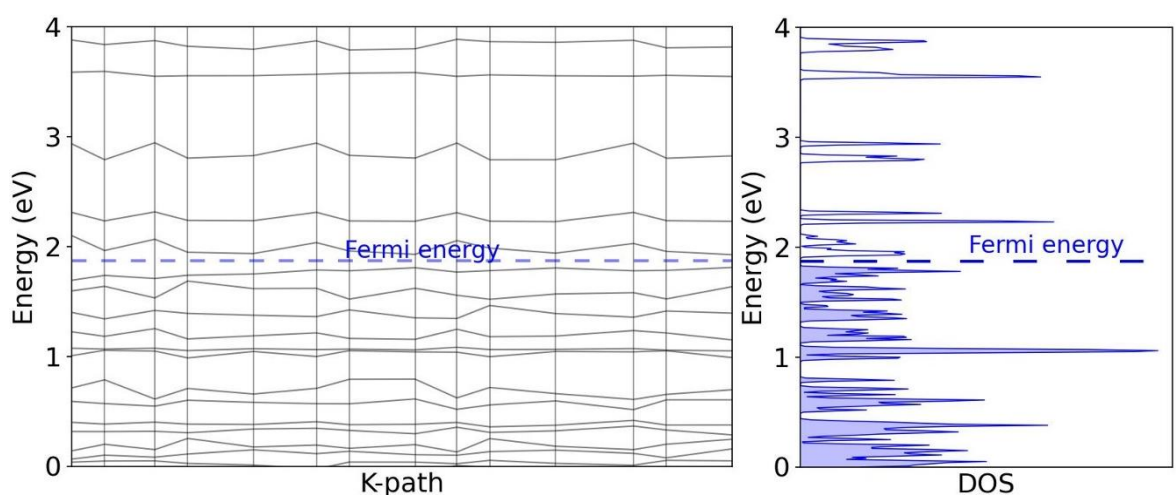

Figure 18. The band structure (left) and DOS (right) of the [4 Zn + Kapton] system. The band gap amounts to 0.12 eV. The Fermi energy is 1.9 eV.

The average electrical conductivity of the soot sample was calculated for 10 randomly selected configurations for each composition, Figure 19. The energy of these configurations differed by no more than 3kT from the energy of the global minimum. The system with polypropylene exhibits the lowest electrical conductivity value, 1.33 ± 0.04 kS m$^{-1}$. The standard errors for these systems are small since the resulting structures are similar. More diverse polymeric structures, such as Kapton, exhibit richer compositions of the soot samples. Therefore, their conductivities also differ substantially from one another. It can be concluded that the PP-based soot obtained during the breakdown is uniform in volume. The value of the electrical conductivity for PET is 2.0 ± 0.2 kS m$^{-1}$, that is, higher than in the case of PP. The electrical conductivity of systems containing PC is $3.0 \pm 0.8$ kS m$^{-1}$. The increased electrical conductivity can be associated with the presence of π-electrons in the molecular fragments containing nanocarbon and heterocycles. In the systems containing the Kapton polymer, the electrical conductivity is the highest one, $7.2 \pm 1.8$ kS m$^{-1}$. This can be associated with a large number of heterocycles, cyclocarbons, and graphene-like structures. In addition, because the structures formed as a result of the experiment are very diverse, the error is about 25%. Recall that higher conductivity is undesirable for our present purposes. The conductivity values shown in Figure 19, calculated using the Kubo-Greenwood theorem, take into account only electronic conductivity. An atomic movement, ionic conductivity, and interfacial contributions were not included. Thus, the received conductivities can be considered indicative only. For example, the calculated conductivity value for aluminum is $0.097 \times 10^7$ S m$^{-1}$, although the experimental value is over an order of magnitude larger, is $3.7 \times 10^7$ S m$^{-1}$.

One can conclude that our KSDFT calculation revealed a slightly underestimated electrical conductivity. However, it can be used for a qualitative assessment to rate more and less desirable soot compositions. The main herein-observed trend is as follows. The electrical conductivity of the soot increases in the following row, PP < PET < PC < Kapton. The computed electrical conductivity expectedly correlates well with the computed band gap. Thus, we can use the computed bandgap to estimate electrical conductivity in further studies to save the simulation resources without compromising the prediction accuracy.

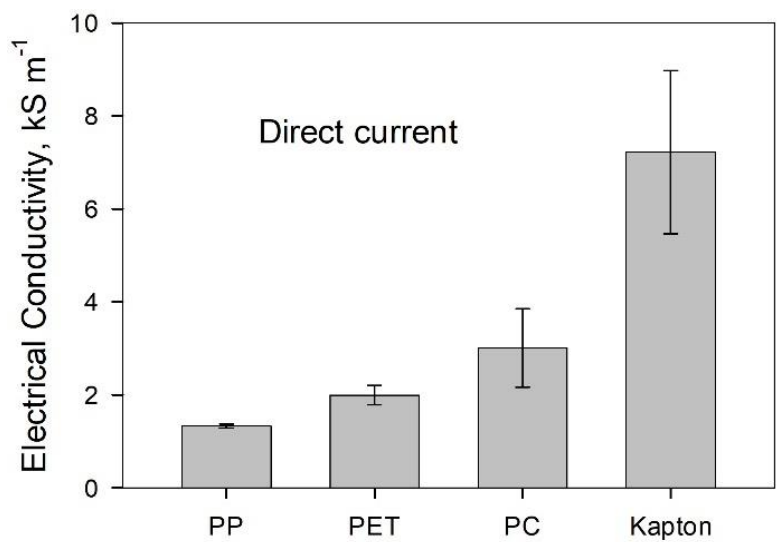

Figure 19. The electrical conductivities of the soot samples consisting of four zinc atoms and four polymers at 293.15 K (degauss) at direct current. The averaging has been carried out over the ten lowest-energy structures of the soot sample.

Figure 20 reveals the effect of the Zn atom fractions in the soot on its electrical conductivity. Note that the content of Zn used in the present work is likely higher than in ordinary expected soot samples. This effect is mostly of fundamental importance. The dependence is reported in the hope of matching some unusual composition of the soot sample after the dielectric breakdown. The electrical conductivity only slightly increases with the Zn content increase. The system containing zero zinc atoms exhibits the lowest conductivity, $(0.97\pm0.07)$ kS m$^{-1}$, compared to the chemical compositions having finite quantities of these atoms. It is an important finding suggesting that the presence of metal atoms is essential to get a somewhat more trustworthy electrical conductivity of the soot even though the soot exhibits a certain semiconductor behavior even in the absence of the Zn atoms.

The highest recorded conductivity amounts to $(1.8 \pm0.3)$ kS m$^{-1}$, which corresponds to the [10 Zn + PP] composition. This is much smaller than that of conductors indicating that the self-healing in the exemplified composition is relatively efficient. The capacitors may continue to work in this condition preventing an immediate short circuit.

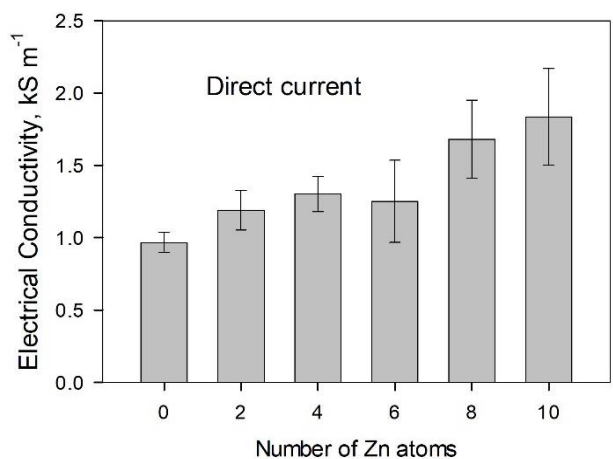

Figure 20. The electrical conductivities of the soot samples containing different fractions of zinc atoms together with the PP polymer at 293.15 K (degauss) at direct current. The averaging has been carried out over the ten lowest-energy structures of the soot sample.

**Conclusions and Final Considerations**

We recapitulate that we herein reported a new theoretical method exploiting electronic-structure calculations at different levels of theory and potential energy surface investigations to assess the feasibility of possible dielectric capacitor setups in the context of self-healing after dielectric breakdowns. The evaluated capacitor setups include (1) the material of the electrode, (2) the material of the insulating polymer, and (3) the ratio of their quantities. We showed that the drastic temperature increase due to the liberated potential energy leads to the formation of numerous gases. The chemical identities and volumes of these gases depend on the insulating polymers employed. For instance, PP gives rise to an impressing amount of molecular hydrogen due to thermodynamic reasons since PP contains the largest fraction of the hydrogen atoms. The formation of gases favors self-healing because gases are obviously non-conductive species. The gases take away atoms, which would otherwise join the solid soot objects, in this way decreasing the size of the adverse semiconductor channel. In turn, the non-volatile products of polymer destruction are carbon-based structures. Such chemical structures exhibit higher electrical conductivities compared to their precursor polymers thanks to the emergence of unsaturated carbon-carbon covalent bonds along with the fused aromatic fragments. The soot samples, which are worse semiconductors than others, are preferable for self-healing.

We evaluated the self-healing potentials of polypropylene, polyethylene terephthalate, polycarbonate, and Kapton with zinc electrodes by computing electrical conductivity, band gaps, and density of states for the unraveled low-energy (most thermodynamically stable) structures within the semiconductor soot samples. These capacitor designs have been widely used in electrotechnical experiments allowing due qualitative comparisons. The calculations showed that the breakdown products of Kapton have the highest conductivities. On the contrary, the products of PP have the lowest conductivities. Based on the overall properties analyzed, the insulating polymers employed with the zinc electrodes display the following decreasing order of self-healing efficiency: PP > PET > PC > Kapton. In terms of electrical conductivity, the trend appeared identical to the general one reported above: PP ($1.33 \pm 0.04$) kS m$^{-1}$ < PET ($2.0 \pm 0.2$) kS m$^{-1}$ < PC ($3.0 \pm 0.8$) kS m$^{-1}$ < Kapton ($7.2 \pm 1.8$) kS m$^{-1}$. In terms of gaseous content, PC outperforms PET but fairly insignificantly: PP (12.3%wt.) > PC (6.4%wt.) > PET (6.2%wt.) > Kapton (5.1wt.%). All the obtained trends are surprisingly similar to the one based on the in-lab electrochemical investigation, PP > PC ≈ PET > Kapton.

Using the new methodology, the work also unveiled an effect of the ratio of the electrode and the polymer. This ratio is an inherent part of the dielectric capacitor design. Furthermore, depending on the localization of an actual dielectric breakdown and the energy involved, the volume of the capacitor burnt may lead to some fluctuations in the soot channel size and composition. We found that the enhanced fraction of zinc in the soot only insignificantly increased the electrical conductivity of the soot. This is because zinc atoms tend to bond to the available non-metallic atoms rather than giving rise to $Zn_n$ chains. If the oxygen atoms were present in the destroyed polymer, Zn preferred to form the fragments of zinc oxide instead of any other structures. We suppose that ZnO decreases the conductivity as it binds high-energy electrons onto lower-energy levels. Indeed, the computed conductivity of the PET-based soot is lower than the one of PC-based soots, the latter containing fewer oxygen atoms.

Were able to correlate specific molecular patterns to the higher values of electrical conductivity. For instance, the structures that are rich in conjugated graphene-like nanocarbons feature higher conductivity values as compared to other soot samples. These chemical formations contain high-energy $sp_2$-hybridized electrons arranged in a nanoscale network. PC and Kapton produce the soot samples hosting such electrons and, indeed, display the highest electrical conductivities recorded herein. Therefore, molecular modeling using a newly elaborated method benefits from a surprisingly convincing qualitative correlation with the experimental data on the life durations of the dielectric capacitors.

The unified simulations were hereby demonstrated to readily rate various capacitor designs according to their vulnerability to a dielectric breakdown occurrence. The lower electrical conductivities of the semiconducting soot bridges correlate with a longer life cycle of the device. The higher molar fractions of the volatile products of the polymer thermal decomposition correlate with a more stable capacitor performance. Indeed, soot samples featuring smaller volumes are less likely to produce conducting bridges that properly unite positive and negative electrodes and lead to a short circuit. In the case that the experimental data on the capacitor stability performance is absent, the simulation results can be validated through the experimental FTIR spectra reflecting the chemical compositions and structures of the soot.

The reported theoretical advance is of high importance for an intensively developed field of dielectric capacitor designs. It allows one to swiftly identify preferable materials and proportions to make a dielectric capacitor. The method relies on the fraction of the volatile by-products formed together with soot and the electrical conductivity of the soot itself. The method directly considers both an insulating polymer and an electrode and their physical and chemical interactions quantum mechanically. The current development is addressed to a wide community of physicists and chemists working at improving the sustainability of dielectric capacitors.

**Supplementary Information**

Figures S1-S2 provide benchmarking data to argue the methodological choices performed in the present work.

**Conflict of interest**

The authors hereby declare no financial interests and professional connections that might bias the interpretations of the obtained results.

**Data availability**

The data supporting this article have been included in the Methodology and the References.

**Acknowledgments**

The research was funded by the Ministry of Science and Higher Education of the Russian Federation under the strategic academic leadership program "Priority 2030" (Agreement 075-15-2024-201 dated 06.02.2024). The results of the work were obtained using computational resources of Peter the Great Saint-Petersburg Polytechnic University Supercomputing Center (www.spbstu.ru).